\documentclass[draftclsnofoot,onecolumn,12pt]{IEEEtran}
\IEEEoverridecommandlockouts
\usepackage{cite}
\usepackage{amsmath,amssymb,amsfonts}
\usepackage{algorithmic}
\usepackage{amsmath} 
\usepackage{graphicx}
\usepackage{textcomp}
\usepackage{xcolor}
\usepackage{amsmath}
\renewcommand{\vec}[1]{\boldsymbol{#1}}
\usepackage{algorithm}
\usepackage{algorithmic}
\usepackage{tcolorbox}
\usepackage{graphicx} 
\usepackage{float} 
\usepackage{stfloats}
\usepackage[section]{placeins}
\usepackage{subfigure}
\usepackage{booktabs}
\usepackage[numbers,sort&compress]{natbib}
\usepackage{amssymb}
\usepackage{setspace}
\usepackage{multirow}
\usepackage{color}
\usepackage{empheq}
\usepackage{pgfplots}
\fontsize{5.0pt}{\baselineskip}\selectfont
\usepackage{epsfig}
\def\BibTeX{{\rm B\kern-.05em{\sc i\kern-.025em b}\kern-.08em
    T\kern-.1667em\lower.7ex\hbox{E}\kern-.125emX}}

\begin{document}
\pgfplotsset{compat=1.14}
\addtolength{\topmargin}{0.04in}
\newlength{\eqnboxwidth}
\setlength{\eqnboxwidth}{\textwidth}
\addtolength{\eqnboxwidth}{-4em}  
\addtolength{\eqnboxwidth}{-.5in} 

\title{The Cram{\'e}r-Rao Bound for  Signal Parameter Estimation from Quantized Data}
\author{Petre~Stoica,~\IEEEmembership{Fellow,~IEEE}, Xiaolei~Shang and Yuanbo~Cheng 
\thanks{This work was supported in part by the Swedish Research Council (VR grants 2017-04610  and 2016-06079), in part by the  National Natural Science Foundation of China under Grant 61771442, and in part by Key Research Program of Frontier Sciences of CAS under Grant QYZDY-SSW-JSC035.}
\thanks{P. Stoica is with the Department of Information Technology, Uppsala University, Uppsala SE-751 05, Sweden (e-mail: ps@it.uu.se).}
\thanks{X. Shang and Y. Cheng are  with the Department of Electronic Engineering and Information
Science, University of Science and Technology of China, Hefei 230027, China
(e-mail: xlshang@mail.ustc.edu.cn and cyb967@mail.ustc.edu.cn).} }
\maketitle
\section*{Introduction and relevance}
Several current ultra-wide band applications, such as millimeter wave radar and communication systems \cite{mo2017channel}\cite{chaoyi}\cite{sun2020mimo}, require high sampling rates and therefore expensive and energy-hungry analog-to-digital converters (ADCs). In applications where cost and power constraints exist, the use of high-precision ADCs is not feasible and the designer must resort to ADCs with coarse quantization. Consequently the interest in the topic of signal parameter estimation from quantized data has increased significantly in recent years.
\par The Cram{\'e}r-Rao bound (CRB) is an important yardstick in any parameter estimation problem. Indeed it lower bounds the variance of any unbiased parameter estimator. Moreover, the CRB is an achievable limit, for instance it is asymptotically attained by the maximum likelihood estimator (under regularity conditions), and thus it is a useful benchmark to which  the accuracy of any parameter estimator  can and should be compared. 
\par A formula for the CRB for signal parameter estimation from real-valued quantized data has been presented in \cite{mezghani2010multiple} but its derivation was somewhat sketchy. The said CRB formula has been extended for instance in \cite{chaoyi} to complex-valued quantized data, but again its derivation was rather sketchy. The special case of binary (1-bit) ADCs and a signal consisting of one sinusoid has been thoroughly analyzed in \cite{host2000effects}. The CRB formula for a binary ADC and a general real-valued signal has been derived, e.g.,  in \cite{gianelli2016one}\cite{gianelli2019one}. 
\par In this lecture note, we  will present a textbook derivation of the CRB for a general signal model and quantizer. We also show that the said CRB monotonically decreases and in the limit converges to the standard CRB (for unquantized data) as the quantization becomes finer and finer. We then consider the special case of binary quantization and present the corresponding CRB. Finally, we show that if the threshold of the binary ADC is allowed to vary in time, then the optimal threshold that minimizes the CRB is the signal itself and  the corresponding CRB is only $\pi/2$ times larger than the standard CRB (a result derived in a different way in \cite{host2000effects} for the special case of one sinusoid in noise). 
\par For the sake of clarity, in the main part of the lecture note we focus on the specific case of normally distributed data, which is most commonly examined in the literature. However, in the appendix titled ``Extensions to general distributions'' we consider data with an arbitrary  distribution and show that the principal results on the CRB derived in the previous sections for the normal distribution can be readily extended to the general case.
\section*{Prerequisites}
While we will try to make this lecture note as self-contained as possible, basic knowledge of statistical signal processing, estimation theory and calculus  will be beneficial for fully understanding it.
\section*{Problem Statement}
Consider the following general model for the noisy measurements of a signal:
\begin{equation}
    y_{n}=s_n(\vec{\theta}) +e_{n}, \quad n=1,\dots,N, \label{eq:model}
\end{equation}
where the index $n$ indicates the sample number, $N$ is the total number of (temporal  or spatial) samples, $s_n(\vec{\theta})$ is the signal model which is a known (differential) function of the unknown parameter vector $\vec{\theta}$, and $e_n$ denotes the noise. We assume that $\lbrace e_n \rbrace_{n=1}^N$ is a sequence of i.i.d. normal random variables with zero mean and known variance $\sigma^2$. We also assume that all  quantities in \eqref{eq:model} are real-valued (the extension to complex-valued measurements is straightforward under the assumption that the normal distribution of the noise is circular, that is  real$(e_n)$ and imaginary$(e_n)$  are independent of each other, see e.g. \cite{chaoyi}). Because $\sigma^2$ is  known, we can divide both sides of \eqref{eq:model}  by $\sigma$ and thus can assume that the noise variance is equal to one, which we will do in what follows to  simplify the notation. The extension to the case of unknown $\sigma$ is not difficult but it leads to more complicated expressions and in order to keep the exposition here as simple as possible, we will not consider it. When there is no risk for confusion, we will omit the dependence of different functions (such as $s_n(\vec{\theta})$) on $\vec{\theta}$ also to simplify the notation and  some of the expressions in the following sections.
\par Consider a quantizer with $b$ bits and $A=2^b$ adjacent intervals defined as follows:
\begin{align}
        I_k&=\left[ l_k, u_k\right), \quad k=1,\dots, A, \nonumber \\
      l_1&=-\infty, \quad u_A=\infty, \quad l_{k+1}=u_k \quad (k=1,\dots,A-1).
\end{align}
When the input to the quantizer lies in $I_k$, the output, denoted $z_k$, is given by
\begin{equation}
    z_k=Q(y) \quad {\rm if} \quad y\in I_k
\end{equation}
and it belongs to a set of size $A$. When the input is $y_n$, the output will be one of the elements in the said set with an index that depends on $n$:
\begin{equation}
    z_{k(n)}=Q(y_n) \quad {\rm for} \quad y_n\in I_{k(n)}.
\end{equation}
The discussion in this lecture note is valid for any desired set of the quantizer output, consequently there is no need to specify the $\lbrace z_k \rbrace$ (for example, in the case of binary ADCs this  set can be $\lbrace -1,1\rbrace$ or $\lbrace 0,1 \rbrace$ etc.).
\par Let  $L(\vec{\theta})$ denote the likelihood function of $\lbrace z_{k(1)},\dots, z_{k(N)}\rbrace$, and let $\vec{J}$ be the Fisher information matrix (FIM) (see \cite{kay1993fundamentals},\cite{stoica1997}):
\begin{equation}
    \vec{J}={E}\left[ \frac{\partial \ln L(\vec{\theta})}{\partial \vec{\theta}} \frac{\partial \ln L(\vec{\theta})}{\partial \vec{\theta}^T}\right]. \label{eq:jfim}
\end{equation}
Because 
\begin{equation}
   {\rm CRB}=\vec{J}^{-1}
\end{equation}
(whenever $\vec{J}$ is nonsingular) our problem is to derive an expression for $\vec{J}$, which we will do in the next section. 
\section*{Solution: FIM for general quantizers}
\subsection{Derivation of the FIM formula}
Under the assumptions made, $\lbrace y_n \rbrace_{n=1}^N$ are independent random variables, therefore so are $\lbrace z_{k(n)}\rbrace_{n=1}^N$. This observation implies that: 
\begin{equation}
    L(\vec{\theta})=p(z_{k(1)},\dots,z_{k(N)})=\prod_{n=1}^N  p(z_{k(n)}), \label{eq:likehood}
\end{equation}
where 
\begin{equation}
    p(z_{k(n)})=p(y_n\in I_{k(n)})=\phi(u_{k(n)}-s_n) -\phi(l_{k(n)}-s_n) \label{eq:pdf}
\end{equation}
with $\phi(x)$ being the cumulative distribution function (cdf) of the normal standard distribution
\begin{equation}
    \phi(x)=\frac{1}{\sqrt{2\pi}}\int_{-\infty}^{x}e^{\frac{-t^2}{2}}dt. \label{eq:cdf}
\end{equation}
From \eqref{eq:likehood} we have that
\begin{equation}
    \frac{\partial \ln L(\vec{\theta})}{\partial \vec{\theta}}=\sum_{n=1}^N \frac{\partial p(z_{k(n)})/\partial \vec{\theta}}{p(z_{k(n)})}. \label{eq:deritive}
\end{equation}
Using \eqref{eq:pdf} and the following standard property of $\phi(x)$,
\begin{equation}
    \phi^{'}(x)=\frac{d\phi(x)}{dx}=\frac{1}{\sqrt{2\pi}} e^{-\frac{x^2}{2}} \quad  ({\rm the \ pdf}) \label{eq:deri},
\end{equation}
we obtain
\begin{equation}
    \frac{\partial p(z_{k(n)})}{\partial \vec{\theta}}=\frac{1}{\sqrt{2\pi}} \left[ e^{-\frac{\left[ l_{k(n)}-s_n\right]^2}{2}}- e^{-\frac{\left[ u_{k(n)}-s_n\right]^2}{2}}\right]\frac{\partial s_n}{\partial \vec{\theta}}, \quad n=1,\dots,N. \label{eq:deri_of_p}
\end{equation}
Because $\lbrace z_{k(n)} \rbrace$ is a sequence of independent random variables, the terms in \eqref{eq:deritive} are independent of each other. Furthermore, the mean of these terms is equal to zero (for $n=1,\dots,N$):
\begin{small}
\begin{align}
    E\left[ \frac{\partial p(z_{k(n)}) / \partial \vec{\theta}}{p(z_{k(n)})} \right]
    &=\sum_{k=1}^A \frac{\partial p(z_{k}) / \partial \vec{\theta}}{p(z_k)} p(z_{k})=\frac{\partial }{\partial \vec{\theta}}\underbrace{\left[\sum_{k=1}^A p(z_k) \right]}_{=1}=0. \label{eq:terms} 
\end{align}
\end{small}
\noindent It follows from \eqref{eq:jfim}, \eqref{eq:deritive} and the discussion above that
\begin{align}
    \vec{J}&=\sum_{n=1}^N\sum_{k=1}^A \frac{\partial p(z_{k}) / \partial \vec{\theta}}{p(z_k)} \frac{\partial p(z_k)/ \partial \vec{\theta}^T}{p(z_k)}p(z_k) \nonumber \\
    &=\sum_{n=1}^N \sum_{k=1}^A \frac{\partial p(z_k) / \partial \vec{\theta} \ \partial p(z_k)/ \partial \vec{\theta}^T}{p(z_k)}. \label{eq:J1}
\end{align}
Inserting \eqref{eq:pdf} and \eqref{eq:deri_of_p} in \eqref{eq:J1} yields the following expression for $\vec{J}$:
\begin{equation}
\framebox{\parbox{\eqnboxwidth}{\centerline{$ \displaystyle
       \vec{J}=\sum_{n=1}^N  \left[ \sum_{k=1}^{A} \frac{\left[ \phi^{'}(u_k-s_n)-\phi^{'}(l_k-s_n)\right]^2}{\phi(u_k-s_n)-\phi(l_k-s_n)}  \right]\frac{\partial s_n}{\partial \vec{\theta}} \frac{\partial s_n}{\partial \vec{\theta}^T} 
$ }
}}
\label{eq:crb}
\end{equation} 
or, more explicitly,
\begin{align}
\framebox{\parbox{\eqnboxwidth}{\centerline{$ \displaystyle
     \vec{J}=\frac{1}{2\pi} \sum_{n=1}^N \left[ \sum_{k=1}^A \frac{\left[ e^{-\frac{(u_k-s_n)^2}{2}} -e^{-\frac{(l_k-s_n)^2}{2}} \right]^2}{\phi(u_k-s_n)-\phi(l_k-s_n)}
    \right]\frac{\partial s_n}{\partial \vec{\theta}}\frac{\partial s_n}{\partial \vec{\theta}^T}.
$}
}}
\label{eq:J}
\end{align}
We note in passing that \eqref{eq:terms} is rarely mentioned in the derivations of the CRB in the literature. However, without \eqref{eq:terms}, the expression for $\vec{J}$ would be more complicated (in particular it would include all cross-terms in the product of \eqref{eq:deritive} with its transpose). 
\subsection{Interval splitting increases the FIM}
Let $\lbrace \tilde{I}_k \rbrace_{k=1}^{\tilde{A}}$ denote a set of intervals obtained by splitting some or all of $\lbrace I_k \rbrace _{k=1}^A$ in smaller subintervals (hence $\tilde{A}> A$) and let $\tilde{\vec{J}}$ be the FIM corresponding to $\lbrace \tilde{I}_k\rbrace$. Intuitively, we would expect that $\tilde{\vec{J}}$  dominates  $\vec{J}$, i.e.
\begin{equation}
\framebox{\parbox{\eqnboxwidth}{\centerline{$ \displaystyle
   \tilde{\vec{J}} \geq \vec{J}
$}
}}
\label{eq:17}
\end{equation} 
in the sense that $(\tilde{\vec{J}}-\vec{J})$ is a positive semi-definite (PSD) matrix.  To prove \eqref{eq:17} we introduce the following notation (we omit the dependence of some of these variables on $k$ and $n$ to simplify the notation):
\begin{align}
     &m_k\in \left[l_k, u_k \right), \nonumber \\
    &a=\phi^{'}(l_k-s_n)-\phi^{'}(m_k-s_n), \nonumber \\
    &b=\phi^{'}(m_k-s_n)-\phi^{'}(u_k-s_n), \nonumber \\
    &\alpha=\phi(u_k-s_n)-\phi(m_k-s_n), \nonumber \\
    &\beta=\phi(m_k-s_n)-\phi(l_k-s_n). 
\end{align}
Below we prove that
\begin{equation}
    \frac{(a+b)^2}{\alpha+\beta} \leq \frac{a^2}{\alpha} + \frac{b^2}{\beta}, \label{eq:ineq_24}
\end{equation}
which clearly implies \eqref{eq:17}. A simple calculation shows that \eqref{eq:ineq_24} is equivalent to the following inequalities:
\begin{align}
    \eqref{eq:ineq_24} &\Leftrightarrow \alpha\beta(a+b)^2 \leq (\alpha+\beta) (\beta a^2+\alpha b^2) \nonumber \\
    &\Leftrightarrow (\alpha b-\beta a)^2 \geq 0.
\end{align}
Because the last inequality above is obviously true, the proof of \eqref{eq:17} is concluded.
\par The optimal splitting point $m_k$ that maximizes the increase of the FIM could be determined by maximizing the right hand side of \eqref{eq:ineq_24}. However, it will depend not only on $k$ but also on $s_n$ and hence it would be of little use from a practical standpoint. This appears to be a general problem for any attempt to optimize the intervals of the quantizer, and some efforts to   circumvent it by assuming that $s_n \approx 0$ (for $n=1,\dots,N$), see e.g. \cite{mezghani2010multiple}, are bound to have only a limited success. Optimizing the intervals $\lbrace I_k\rbrace$ by maximizing the FIM is an interesting research problem that awaits a practically useful general solution (an efficient global solver for this interval design problem will be presented in the forthcoming paper \cite{interval_opt}). 
\subsection{Upper and lower bounds on the FIM}
Let 
\begin{equation}
    \rho=\sum_{k=1}^A \frac{\left[ \phi^{'}(\tilde{u}_k)- \phi^{'}(\tilde{l}_k)\right]^2}{\phi(\tilde{u}_k)-\phi(\tilde{l}_k)},
\end{equation}
where $\phi(x)$ and  $\phi^{'}(x)$ are as defined in \eqref{eq:cdf}  and \eqref{eq:deri}, and 
\begin{equation}
    \tilde{u}_k=u_k-s_n, \quad \tilde{l}_k=l_k-s_n \ (\tilde{u}_k > \tilde{l}_k) \label{eq:22}
\end{equation}
(we omit the dependence of $\rho$, $\tilde{u}_k$ and $\tilde{l}_k$ on $n$ to simplify the notation).
\par In comparison with \eqref{eq:crb}, the standard CRB (for unquantized data) has a simpler expression that corresponds to setting $\rho\equiv 1$ in \eqref{eq:crb}:
\begin{equation}
     \vec{J}_0=\sum_{n=1}^N \frac{\partial s_n}{\partial \vec{\theta}} \frac{\partial s_n}{\partial \vec{\theta}^T} 
     \label{eq:J0}   
\end{equation}
(see, e.g., \cite{kay1993fundamentals}\cite{stoica1997}). Because the output of the quantizer  provides less ``information'' about the signal than the unquantized data $\lbrace y_n \rbrace$, we expect that
\begin{equation}
\framebox{\parbox{\eqnboxwidth}{\centerline{$ \displaystyle
    \vec{J} \leq \vec{J}_0. 
$}
}}
\label{eq:24}
\end{equation}
Proving that \eqref{eq:24} indeed holds is an interesting exercise that we undertake in what follows. First we note that
\begin{align}
    \left[ \phi^{'}(\tilde{u}_k)-\phi^{'}(\tilde{l}_k)\right]^2&=\left[ \int_{\tilde{l}_k}^{\tilde{u}_k} \phi^{''}(x) \ dx\right]^2=\left[ \int_{\tilde{l}_k}^{\tilde{u}_k}  \frac{\phi^{''}(x)}{\sqrt{\phi^{'}(x)}}\sqrt{\phi^{'}(x)} \ dx\right]^2 \nonumber \\
    & \leq \left[ \int_{\tilde{l}_k}^{\tilde{u}_k}\phi^{'}(x) \ dx\right] \left[\int_{\tilde{l}_k}^{\tilde{u}_k}  \frac{[\phi^{''}(x)]^2}{\phi^{'}(x)} \ dx \right] \nonumber \\
    &=\left[\phi(\tilde{u}_k)-\phi(\tilde{l}_k) \right]\int_{\tilde{l}_k}^{\tilde{u}_k} \frac{[\phi^{''}(x)]^2}{\phi^{'}(x)} \ dx. \label{eq:25}
\end{align}
The inequality in \eqref{eq:25} follows from the Cauchy-Schwarz inequality for integrals. Using \eqref{eq:25} in the definition of $\rho$ yields the following inequality:
\begin{equation}
    \rho \leq \sum_{k=1}^A \int_{\tilde{l}_k}^{\tilde{u}_k} \frac{[\phi^{''}(x)]^2}{\phi^{'}(x)}dx=\int_{-\infty}^{\infty} \frac{[\phi^{''}(x)]^2}{\phi^{'}(x)}dx=
    \frac{1}{\sqrt{2\pi}}\int_{-\infty}^{\infty} x^2 e^{-\frac{x^2}{2}}dx. \label{eq:26}
\end{equation}
However the last expression in \eqref{eq:26} is nothing but the second-order moment of a normal random variable with zero mean and variance equal to 1, therefore:
\begin{equation}
    \rho \leq 1 \label{eq:27}
\end{equation}
and \eqref{eq:24} is proved. \\
\indent It follows from \eqref{eq:27} and the analysis in the previous subsection that, as the splitting of the intervals becomes finer, the corresponding values of $\rho$  form a monotonically increasing sequence of numbers that are bounded above by 1. The monotonic convergence theorem (see, e.g., \cite{rudin1976principles}) then implies that this sequence converges. In the appendix titled ``The convergence of $\rho$''  we prove that 
\begin{equation}
    \rho \rightarrow 1  \label{eq:28}
\end{equation}
and thus 
\begin{equation}
\framebox{\parbox{\eqnboxwidth}{\centerline{$ \displaystyle
    \vec{J} \to \vec{J}_0 
    $}
    }}
    \label{eq:converg}
\end{equation} 
as $A\to \infty$ and the quantization becomes infinitely fine. 
\par For a finite $A$, however, in general $\vec{J}$ will be strictly less than $\vec{J}_0$. In the last part of this subsection, we will derive a lower bound on $\vec{J}$ that can be used to determine how large the difference ($\vec{J}_0-\vec{J}$) can be. To that end, we will use the notation introduced in \eqref{eq:22}, and the following result that is a consequence of the Cauchy-Schwarz inequality:
\begin{equation}
    (\sum_{k=1}^A \left| a_k\right|)^2=(\sum_{k=1}^A\frac{\left| a_k\right|}{\sqrt{b_k}}\sqrt{b}_k)^2 \leq (\sum_{k=1}^A \frac{a_k^2}{b_k}) (\sum_{k=1}^A b_k) \quad {\rm for} \quad \lbrace b_k >0 \rbrace_{k=1}^A  \nonumber
\end{equation}
which implies that:
\begin{equation}
    \sum_{k=1}^A \frac{a_k^2}{b_k} \geq (\sum_{k=1}^A \left|a_k \right|)^2 \ {\rm for} \ \lbrace b_k >0 \rbrace_{k=1}^A  \ {\rm and} \ \sum_{k=1}^A b_k=1. \label{eq:30}
\end{equation}
Using the following definitions,
\begin{align}
      \left| a_k \right|&=\left| e^{-\frac{\tilde{u}_k^2}{2}} -e^{-\frac{\tilde{l}_k^2}{2}} \right| \nonumber  \\
      b_k&=\phi(\tilde{u}_k)-\phi({\tilde{l}_k})
\end{align}
in \eqref{eq:30} we obtain the inequality:
\begin{equation}
    \rho=\frac{1}{2\pi} \sum_{k=1}^A \frac{a_k^2}{b_k} \geq \frac{1}{2\pi} \left[ \sum_{k=1}^A |a_k|\right]^2. \label{eq:32}
\end{equation}
Now, let $d$ be such that $s_n \in \left[l_d, u_d \right)$ or equivalently
\begin{equation}
    \left\{\begin{aligned}
        \tilde{l}_k\leq 0 \quad  &{\rm  for} \quad  k=1,\dots,d \\
        \tilde{l}_k > 0  \quad & {\rm for} \quad k=d+1,\dots, A \label{eq:d33}\\ 
\end{aligned}
\right. 
\end{equation}
(the dependence of $d$ on $n$ is omitted to simplify the notation; this dependence on $n$ will be reinstated when it becomes important). Using the above definition of $d$ it is straightforward to check that:
\begin{align}
    \eta  \triangleq \frac{1}{2} \sum_{k=1}^A |a_k|&=\frac{1}{2} \left[ ( e^{-\frac{\tilde{u}_1^2}{2}} -e^{-\frac{\tilde{l}_1^2}{2}}) +  (e^{-\frac{\tilde{u}_2^2}{2}} -e^{-\frac{\tilde{l}_2^2}{2}}) + \dots + 
    \left| e^{-\frac{\tilde{u}_d^2}{2}} -e^{-\frac{\tilde{l}_d^2}{2}} \right| \right. \nonumber \\
     & \left. \quad + (e^{-\frac{\tilde{l}_{d+1}^2}{2}} -e^{-\frac{\tilde{u}_{d+1}^2}{2}})
    + (e^{-\frac{\tilde{l}_{d+2}^2}{2}} -e^{-\frac{\tilde{u}_{d+2}^2}{2}})
    + \dots + (e^{-\frac{\tilde{l}_{A}^2}{2}} -e^{-\frac{\tilde{u}_{A}^2}{2}}) \right] \nonumber \\
    &=\left\{
    \begin{aligned}
    e^{-\frac{\tilde{u}_d^2}{2}} &\quad {\rm if} \quad \tilde{u}_d^2 \leq \tilde{l}_d^2 \\
    e^{-\frac{\tilde{l}_d^2}{2}} &\quad {\rm if} \quad \tilde{u}_d^2 > \tilde{l}_d^2. 
    \end{aligned}
    \right. \label{eq:34}
\end{align}
Combining \eqref{eq:32} and \eqref{eq:34} yields the following lower bound on $\vec{J}$:
\begin{equation}
\framebox{\parbox{\eqnboxwidth}{\centerline{$ \displaystyle
    \vec{J}\geq \frac{2}{\pi} \sum_{n=1}^N \eta^2(n) \frac{\partial s_n}{\partial \vec{\theta}} \frac{\partial s_n}{\partial \vec{\theta}^T}. 
    $}
    }}
    \label{eq:35}
\end{equation}
As already mentioned, $s_n$ lies in the interval $\left[l_{d(n)}, u_{d(n)} \right)$.
The smaller this interval the larger $\eta(n)$, and thus the larger the lower bound in \eqref{eq:35}. 
\section*{Solution: FIM for a binary quantizer}
\subsection{Derivation of the FIM formula}
For $b=1$, we have $l_1=-\infty$, $u_1=l_2=0$, and $u_2=\infty$. Consequently, 
\begin{equation}
    \vec{J}_{1}=\frac{1}{2\pi}\sum_{n=1}^N \left[ \frac{e^{-s_n^2}}{\phi(-s_n)}+ \frac{e^{-s_n^2}}{1-\phi(-s_n)}\right]\frac{\partial s_n}{\partial \vec{\theta}} \frac{\partial s_n}{\partial \vec{\theta}^T}
\end{equation}
or equivalently
\begin{equation}
\framebox{\parbox{\eqnboxwidth}{\centerline{$ \displaystyle
    \vec{J}_{1}=\frac{1}{2\pi} \sum_{n=1}^N \frac{e^{-s_n^2}}{\phi(s_n)\phi(-s_n)}\frac{\partial s_n}{\partial \vec{\theta}} \frac{\partial s_n}{\partial \vec{\theta}^T}. 
    $}
    }}\label{eq:J_onebit}
\end{equation}
A proof of \eqref{eq:J_onebit} was presented in \cite{gianelli2019one} but it relied on a specific choice of the set $\lbrace z_k \rbrace$ (namely $z_1=-1$ and $z_2=1$), while the above derivation holds for any choice of  $\lbrace z_k \rbrace$. 
\subsection{Upper and lower bounds on the FIM}
The analysis in \cite{host2000effects} of the special case of one sinusoid in noise found out that
\begin{equation}
\framebox{\parbox{\eqnboxwidth}{\centerline{$ \displaystyle
    \vec{J}_1 \leq \frac{2}{\pi} \vec{J}_0
    $}
    }}
    \label{eq:J_one_upper}
\end{equation}
where $\vec{J}_0$ is the standard FIM, see \eqref{eq:J0}. We will show that \eqref{eq:J_one_upper} also holds for the general signal model considered here, but before doing so we should like to determine if the upper bound in \eqref{eq:J_one_upper} is achievable. To provide an answer to this question, we will consider quantizers with time-varying thresholds, e.g., like in \cite{gianelli2016one}.  In other words, instead of comparing $y_n$ to zero, we compare it to a known threshold denoted $h_n$. The only modification of the expression of $\vec{J}_1$ that this change of threshold entails is that the first factor  in   \eqref{eq:J_onebit} should be replaced by:
\begin{equation}
\rho_1=\frac{e^{-(s_n-h_n)^2}}{\phi(s_n-h_n)\phi(h_n-s_n)}. \label{eq:39}
\end{equation}
Clearly for $h_n=s_n$, we get $\rho_1=4$ and thus the bound in \eqref{eq:J_one_upper} is attained. Consequently the optimal threshold that minimizes the ${\rm CRB}=\vec{J}_1^{-1}$ is the signal itself. While $\lbrace s_n \rbrace$ of course is unknown (at least initially), this choice of threshold could be implemented sequentially in $n$ as more data are collected and better estimates of $\lbrace s_n \rbrace$ become available. However, we should keep in mind  the fact that generating a time-varying threshold $\lbrace h_n\rbrace$ will  require DACs, which can be as expensive and energy-hungry as the ADCs, and therefore may also be limited in the number of bits they can use. It is an interesting question if a simple binary ADC combined with a DAC that uses $b$ bits to generate an approximation of the optimal threshold $h_n=\hat{s}_n$ (from the most recent estimate $\hat{s}_n$ of $s_n$) yields better (or worse) performance than using only one $b$-bit ADC (see the discussion at the end of this section and the numerical evaluation section for a partial answer). 
\par We now return to the inequality in \eqref{eq:J_one_upper}  and present a proof of it based on elementary arguments. Clearly \eqref{eq:J_one_upper} follows if we can show that:
\begin{equation}
    \frac{e^{-s^2}}{\phi(s)\phi(-s)} \leq 4, \forall s 
\end{equation}
or, equivalently,
\begin{equation}
    f(s)\triangleq 4\phi(s)\phi(-s) -e^{-s^2} \geq 0, \forall s.  \label{eq:41}
\end{equation}
For later use note that
\begin{equation}
    f(-\infty)=0, \quad f(0)=0, \quad {\rm and}\quad  f(\infty)=0. \label{eq:valuef}
\end{equation}
A simple calculation yields
\begin{align}
    f^{'}(s)&=\frac{4}{\sqrt{2\pi}} \left[ e^{-\frac{s^2}{2}}\phi(-s)-e^{-\frac{s^2}{2}}\phi(s) \right] + 2 s e^{-s^2} \nonumber  \\
    &=2e^{-\frac{s^2}{2}}g(s)
\end{align}
where 
\begin{equation}
    g(s)=\frac{2}{\sqrt{2\pi}} \left[\phi(-s)-\phi(s) \right] + s e^{-\frac{s^2}{2}}.
\end{equation}
Clearly 
\begin{equation}
    f^{'}(s)=0 \Leftrightarrow g(s)=0 \ {\rm and} \ {\rm sign}[ f^{'}(s)]={\rm sign}\left[ g(s) \right], \ \forall |s| < \infty.  \label{eq:signg}
\end{equation}
Also, 
\begin{equation}
    g(-\infty)=\sqrt{\frac{2}{\pi}},  \  g(0)=0, \ {\rm and} \ g(\infty)=-\sqrt{\frac{2}{\pi}}.  \label{eq:goned}
\end{equation}
Next we note that
\begin{equation}
    g^{'}(s)=\frac{2}{2\pi} \left[-e^{-\frac{s^2}{2}} -e^{-\frac{s^2}{2}}\right]+ e^{-\frac{s^2}{2}} -s^2e^{-\frac{s^2}{2}}=e^{-\frac{s^2}{2}}\left[(1-\frac{2}{\pi})-s^2 \right]. \label{eq:gd}
\end{equation}
Let 
\begin{equation}
    s_1=-\sqrt{1-2/\pi}, \ s_2=\sqrt{1-2/\pi}
\end{equation}
denote the finite roots of \eqref{eq:gd} and observe that
\begin{equation}
    g^{''}(s_i)=-2s_ie^{-\frac{s_i^2}{2}}, \quad (i=1,2) 
\end{equation}
which implies that
\begin{equation}
    s_1={\rm min \ point \ of} \ g(s), \  s_2={\rm max \ point \ of} \ g(s). \label{eq:minmax}
\end{equation}
Combining the facts shown above, see Fig. 1 and its caption, shows that $f(s)$ has a minimum at $s=0$ (where $f(0)=0$) and two maximum points, and therefore it satisfies 
\eqref{eq:41}. With this observation, the proof is concluded.
\begin{figure}[htb]
\centering
\label{fig:fig1}
\begin{minipage}[t]{0.38\linewidth}
\centering
\centerline{\epsfig{figure=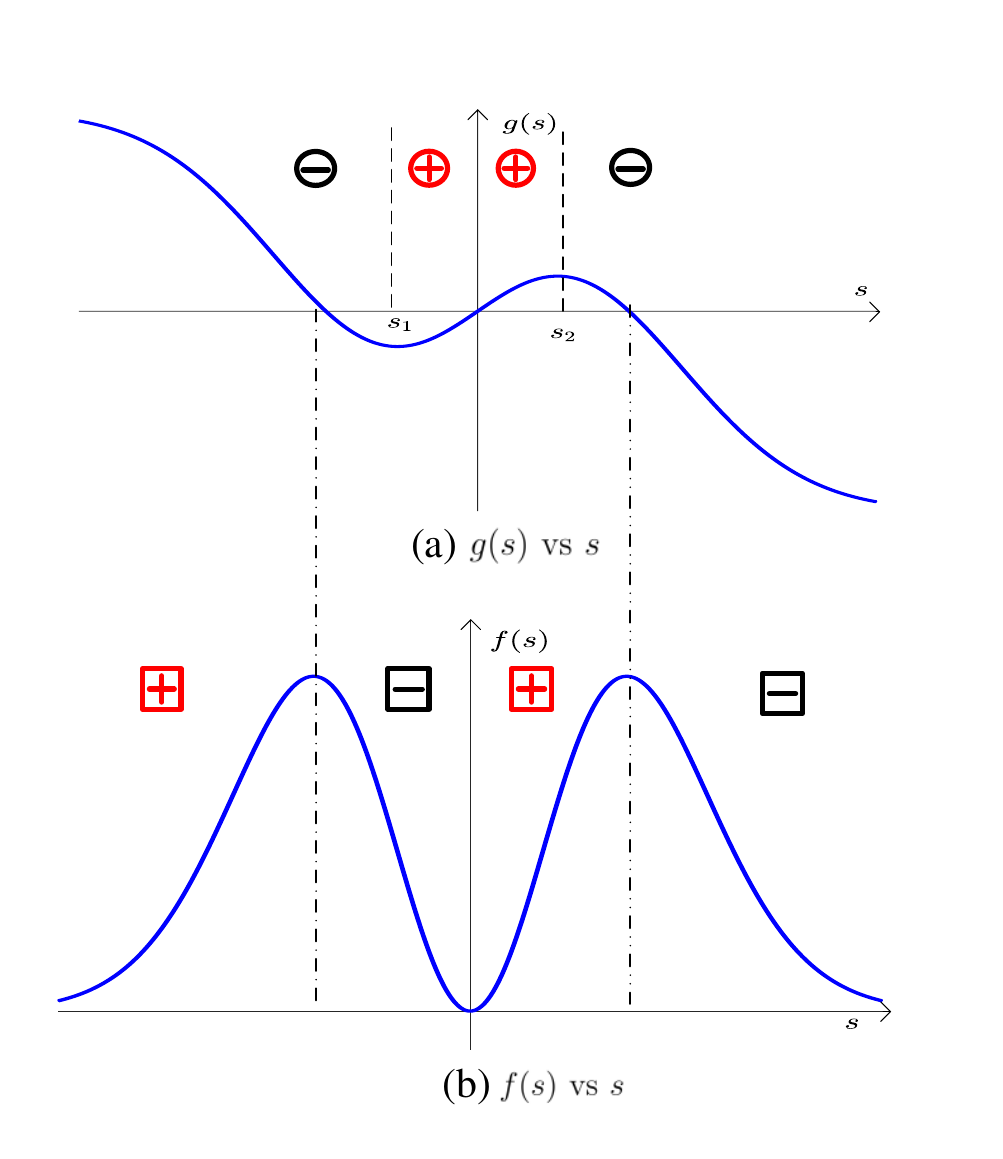,width=8cm}}
\end{minipage}
\caption{ (a) ${\rm sign}[g^{'}(s)]$ (indicated by circles) is positive in $[s_1,s_2]$ and negative elsewhere (see \eqref{eq:gd}). This observation along with the fact that $s_1$ is a minimum of $g(s)$ and $s_2$ a maximum (see \eqref{eq:minmax}), and that $g(s)$ takes on the values in \eqref{eq:goned} lead to the plot of $g(s)$ in this part of the figure. (b) Using the fact that ${\rm sign}[f^{'}(s)]={\rm sign}[g(s)]$ (see \eqref{eq:signg}) we get the variation of ${\rm sign}[f^{'}(s)]$ indicated by squares. This variation along with the values of $f(s)$ in \eqref{eq:valuef} and the fact that $f^{'}(s)=0\Leftrightarrow g(s)=0$ for any $|s|<\infty$ (see \eqref{eq:signg}) lead to the plot of $f(s)$ in this subfigure, which shows that $f(s)\geq 0  \ \forall s$.}
\end{figure}
\par At the end of this subsection, we present  a lower bound on $\vec{J}_1$ and compare it with the lower bound in \eqref{eq:35} on $\vec{J}$. In the case of a binary quantizer $d$ in  \eqref{eq:d33} is either 1 or 2 (depending on whether $s_n <0$ or $s_n\geq 0$). In either case $\eta$ in \eqref{eq:34} is given by $e^{-\frac{s_n^2}{2}}$  and hence \eqref{eq:35}  reduces to: 
\begin{equation}
\framebox{\parbox{\eqnboxwidth}{\centerline{$ \displaystyle
    \vec{J}_1 \geq \frac{2}{\pi} \sum_{n=1}^N e^{-s_n^2} \frac{\partial s_n}{\partial \vec{\theta}}\frac{\partial s_n}{\partial \vec{\theta}^T}.
    $}
    }}
    \label{b} 
\end{equation}
The lower bound in \eqref{b} will typically be significantly smaller than that in \eqref{eq:35}, in agreement with the fact that the estimation performance corresponding to a binary quantizer is inferior to that of a quantizer using more bits (see \eqref{eq:17}). On the other hand, if $e^{-s_n^2}$ in \eqref{b} is replaced by $e^{-(h_n-s_n)^2}$ (as in \eqref{eq:39}), with the threshold $h_n=\hat{s}_n$ generated using a DAC with $b$ bits, then the lower bounds in \eqref{eq:35} and \eqref{b} become quite similar to one another. In such a case, if $b$ is sufficiently large (for example $b\geq 4$) the two bounds are well approximated by $\frac{2}{\pi}\vec{J}_0$. For the binary ADC, this matrix is also an upper bound, see \eqref{eq:J_one_upper}, and hence it corresponds to the apex performance of this quantizer. However, the FIM for an ADC with $b>1$ can  be larger than $\frac{2}{\pi}\vec{J}_0$, which means that the estimation performance afforded by a $b$-bit ADC can in principle be  better than that of a 1-bit ADC even if the latter uses an optimal threshold generated by a $b$-bit DAC. (see the next section for an illustration).
\par We remark in this context that a tighter upper bound on $\vec{J}$ than $\vec{J}\leq \vec{J}_0$ (similar to the bound in \eqref{eq:J_one_upper}) is not available. Derivation of such a bound on $\vec{J}$, which would  generalize \eqref{eq:J_one_upper} to the case of $b>1$, is  an open research problem. However, we must note that, while such a bound would be theoretically interesting, its practical usefulness would be limited. Indeed, $\rho$ is quite close to one (and hence $\vec{J}$ to $\vec{J}_0$) even for relatively small values of $b$, such as $b=3$ or $b=4$ (for which $\max\limits_{\lbrace I_k\rbrace}\rho$  is 0.97 and 0.99, respectively), and  for all practical purposes $\max\limits_{\lbrace I_k\rbrace}\rho$ becomes indistinguishable from  one as $b$ increases. Consequently, an efficient and reliable algorithm for computing the optimum $\lbrace I_k \rbrace$ that maximize $\rho$ appears to be a more useful practical contribution than an upper bound on $\rho$ (see the future paper \cite{interval_opt} for such an algorithm).
\section*{Numerical Example}
\noindent We consider a signal comprising two sinusoids:
\begin{equation}
    s_n(\vec{\theta})=a_1 \sin(\omega_1 n+\varphi_1) +a_2 \sin{(\omega_2n+\varphi_2)}, \quad n=1,2,\dots, N,
\end{equation}
where
\begin{align}
    N&=100, \ {\rm or} \ 512, \nonumber \\
    \vec{\theta}&=\left[
    \begin{matrix}
    a_1 & a_2 & \omega_1 & \omega_2 & \varphi_1 &\varphi_2 
    \end{matrix}\right]^T, \nonumber \\
    \omega_1&=0.25, \quad  \omega_2=0.4, \nonumber \\
    a_1&=1, \quad a_2=1/r, \nonumber \\
    \varphi_1&=\pi/3, \quad \varphi_2=\pi/4.
\end{align}
We vary $r$ from 1 to 200. We also vary the noise variance to maintain the same ${\rm SNR}_2=0 \ {\rm dB}$ for the weaker sinusoid, for all values of $r$, where:
\begin{equation}
    {\rm SNR}_2=\frac{a_2^2}{2\sigma^2}.
\end{equation}
The intervals $\lbrace I_k \rbrace$ are chosen as suggested in \cite{max1960quantizing} for a Lloyd-Max quantizer. In the case of $b=4$, considered here, and for a signal with unit power these intervals are given by:
\begin{align}
    \lbrace &-\infty, -2.401, -1.844, -1.437, -1.099, -0.7996, -0.5224, -0.2582,  \nonumber \\ 
    & 0.0000, 0.2582, 0.5224, 0.7996, 1.099, 1.437,  1.844, 2.401, \infty \rbrace. \nonumber
\end{align}
\par
\begin{figure}[htb]
\centering
\subfigure[CRB($\omega_1$),\ $N=100$]{
\label{fig:crbw1_100} 
\begin{minipage}[t]{0.38\linewidth}
\centering
\centerline{\epsfig{figure=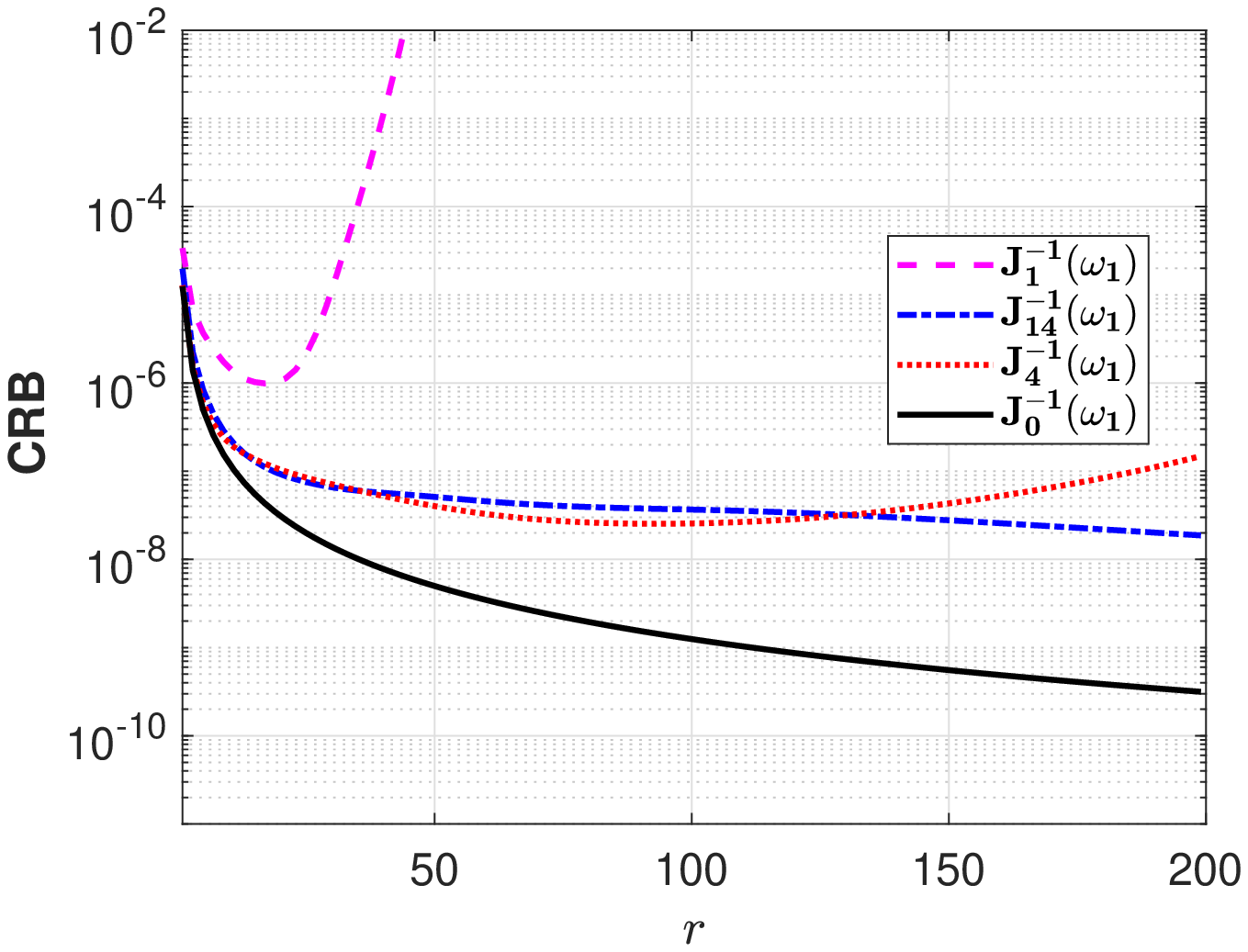,width=7cm}}
\end{minipage}} 
\subfigure[CRB($\omega_2$),\ $N=100$]{
\label{fig:crbw2_100}
\begin{minipage}[t]{0.38\linewidth}
\centering
\centerline{\epsfig{figure=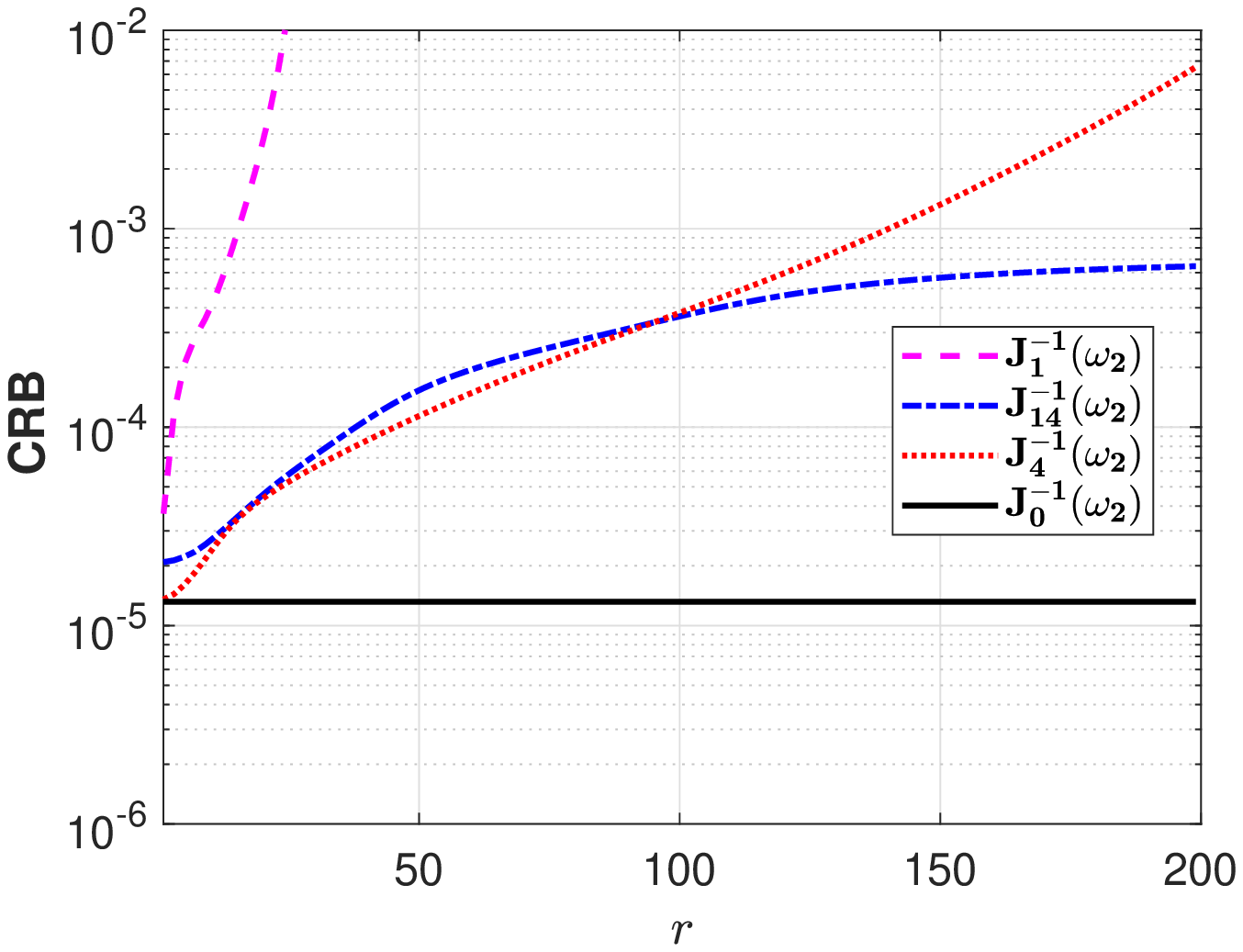,width=7cm}}
\end{minipage}}
\caption{CRB versus $r$ for $\omega_1$ (a) and $\omega_2$ (b). $\vec{J}_0^{-1}$ = standard CRB (for unquantized data), $\vec{J}_{1}^{-1}$=CRB for 1-bit ADC, $\vec{J}^{-1}_4$=CRB for 4-bit ADC, and  $\vec{J}_{14}^{-1}$=CRB for 1-bit ADC using the optimal threshold $h_n=s_n$ generated via a 4-bit DAC. ($N=100$).}
\label{fig:simu_100}
\end{figure}

\begin{figure}[htb]
\centering
\subfigure[CRB($\omega_1$),\ $N=512$]{
\label{fig:crbw1_512}
\begin{minipage}[t]{0.38\linewidth}
\centering
\centerline{\epsfig{figure=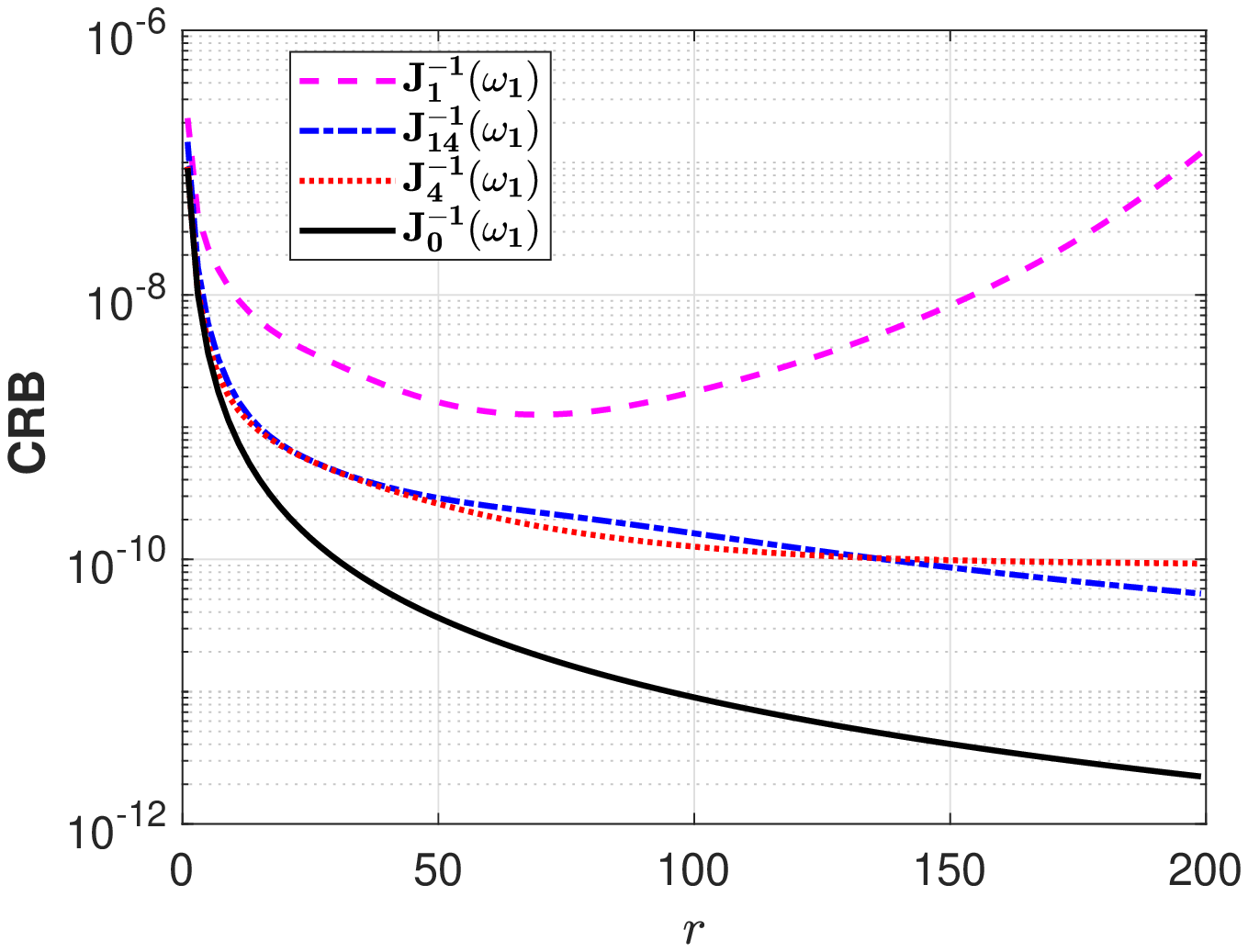,width=7cm}}
\end{minipage}} 
\subfigure[CRB($\omega_2$),\ $N=512$]{
\label{fig:crbw2_512}
\begin{minipage}[t]{0.38\linewidth}
\centering
\centerline{\epsfig{figure=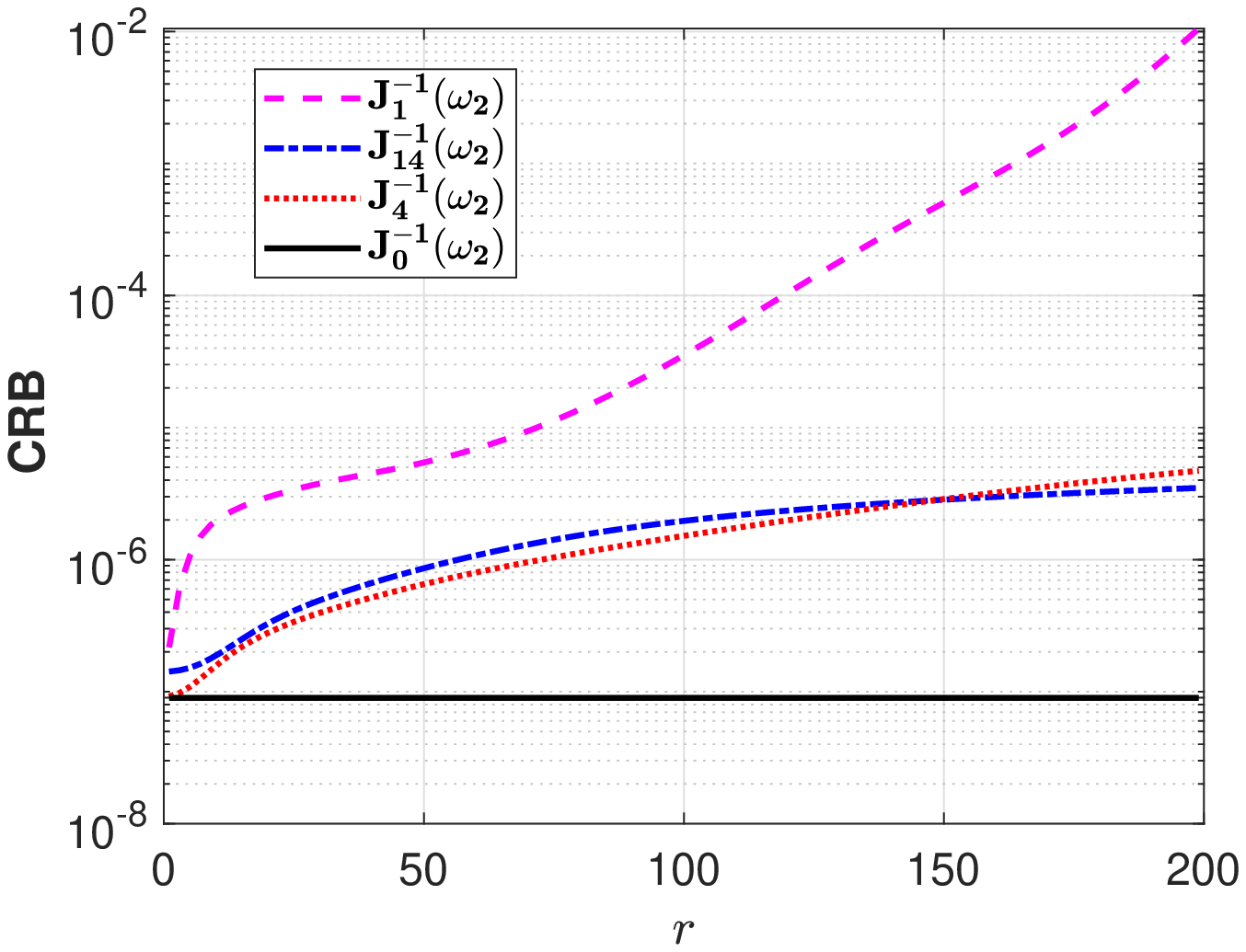,width=7cm}}
\end{minipage}}
\caption{CRB versus $r$ for $\omega_1$ (a) and $\omega_2$ (b). $\vec{J}_0^{-1}$ = standard CRB (for unquantized data), $\vec{J}_{1}^{-1}$=CRB for 1-bit ADC, $\vec{J}^{-1}_4$=CRB for 4-bit ADC, and  $\vec{J}_{14}^{-1}$=CRB for 1-bit ADC using the optimal threshold $h_n=s_n$ generated via a 4-bit DAC. ($N=512$).}
\label{fig:simu_512}
\end{figure}

 Figs. \ref{fig:simu_100} and \ref{fig:simu_512} show four CRBs for $\omega_1$ and $\omega_2$, see the explanation in the figure's caption. The standard CRB for $\omega_1$, $\vec{J}_0^{-1}(\omega_1)$, decreases with $r$ because $\sigma^2$ decreases as  $r$ increases (as explained above) and hence  ${\rm SNR}_1$ increases with $r$. For the second sinusoid, $\vec{J}^{-1}_0(\omega_2)$ is constant as $r$ varies because ${\rm SNR}_2$ is the same for all values of $r$. 
\par The 1-bit CRB, $\vec{J}_1^{-1}$, is the largest of the considered bounds, which is the price paid for the simplicity of the binary quantizer. Moreover the degradation of $\vec{J}^{-1}_1$  as $r$ increases is significant and can reach  unacceptable levels if $N$ is not large enough. Both $\vec{J}^{-1}_4$ and  $\vec{J}_{14}^{-1}$ are much smaller than $\vec{J}_1^{-1}$, and both degrade more gracefully in the case of $\omega_2$, as $r$ increases, and remain relatively close to $\vec{J}^{-1}_0$. As one can see from Figs. \ref{fig:simu_100} and \ref{fig:simu_512}, typically,  $\vec{J}_{4}^{-1}$ is not far from than $\vec{J}_{14}^{-1}$, despite the fact that the latter is based on  information about the signal. 
\par The bowl-shaped plot of $\vec{J}_1^{-1}(\omega_1)$ in Figs. \ref{fig:crbw1_100} and \ref{fig:crbw1_512} can be explained as follows. The second sinusoid acts as (unknown) jitter on the first sinusoid and it is a well-known fact that jitter can improve the accuracy of parameter estimation from binary data. The optimal power of the jitter (or the corresponding value of $r$), which minimizes $\vec{J}_1^{-1}(\omega_1)$, is problem dependent. A similar effect occurs due to the noise (we remind the reader that  the noise variance decreases as $r$ increases), which acts as dither on the signal. This jittering/dithering effect causes  the decrease of the $\vec{J}_1^{-1}(\omega_1)$ in both  Figs. \ref{fig:crbw1_100} and \ref{fig:crbw1_512}, but as $r$ increases the said effect vanishes and $\vec{J}_1^{-1}(\omega_1)$ starts to increase. If we continue to increase $r$ beyond the range in the figure, $\vec{J}_1^{-1}(\omega_1)$ increases without bound, a fact which indicates that parameter identifiability has been lost.  
\par Note that the curves of $\vec{J}_{14}^{-1}(\omega_1)$ and $\vec{J}_4^{-1}(\omega_1)$ are much more stable for $r$ in the range considered in the figure. However if $r$ is increased beyond a certain level (admittedly too large for being of practical interest)  then $\vec{J}_4^{-1}(\omega_1)$ also starts to increase and eventually this scheme also loses parameter identifiability. In  contrast to this in our experiments $\vec{J}_{14}^{-1}(\omega_1)$ has continued to decrease even when we let $r$ take on quite large values. 
\par In sum, the main findings of this  numerical evaluation are:
\begin{itemize}
\item a 4-bit quantizer appears to offer satisfactory estimation performance in a wide range of situations, including cases with signal components of rather different powers  (but excluding cases in which the signal is almost noise-free, a situation that obviously is unlikely to occur in practical applications). 
\item the use of 1-bit ADC along with a $4$-bit DAC for threshold generation does not appear to offer  any advantage, either in hardware or  estimation performance, over employing a single 4-bit ADC, even when the threshold generation block had full information about the signal  (unless there is very little noise in the data, which is a case of theoretical rather than practical interest).
\item when a 1-bit quantizer is the only feasible option, the user should keep in mind that the estimation performance offered by this quantizer can degrade very quickly as the dynamic range of the signal components increases, unless $N$ is sufficiently large. A study of the CRB using the formulas presented in this lecture note can offer guidelines for the cases in which the use of a 1-bit quantizer can be a viable solution. 
\end{itemize}
\section*{What we have learned}
The clear derivation and simple form of the CRB expression for signal parameter estimation from quantized  data, presented in this lecture note, will hopefully encourage the more frequent  use of the CRB in applications where coarse-quantization ADCs are the only feasible choice, such as in ultra-wide band radar and communication systems or in large sensor networks. Proving that the said CRB monotonically decreases and in the limit converges to the standard CRB (for the unquantized data) as the quantization becomes finer and finer can be viewed as an interesting exercise  in statistical signal processing and calculus; and so can the analysis leading to the upper bound on the CRB as well as  the achievable lower bound on the CRB associated with 1-bit ADCs. The discussion on the optimal threshold in applications using 1-bit ADCs, as well as the performance comparison  between a system employing  a $b$-bit ADC and another system  using a 1-bit ADC in combination with a $b$-bit DAC (for threshold generation), also are of potential interest for applications. 
\par Finally, this lecture note has illustrated the fact that, as the number of bits of the quantizers decreases, the estimation accuracy for the small components degrades much more than  for the large components of the signal  (as intuitively expected: when $A=2^b$ is small, the few available intervals of the quantizer are normally chosen sufficiently large to cover  the variation of the large components in the signal, and hence most information about the small components is lost in the quantized data. ). This dynamic range problem also appears to be practically relevant and worth of further study.

\appendix
\section*{A. \upshape  The convergence of $\rho$} 
We  will use the following notation:
\begin{equation}
     \mu_k=\phi(\tilde{l}_k), \ \nu_{k}=\phi(\tilde{u}_k),
\end{equation}
and note that 
\begin{equation}
    0=\mu_1 < \nu_1 \dots \mu_A < \nu_A=1 \ \left({\rm i.e.,} \ [0,1)=\sqcup_{k=1}^A [\mu_k,\nu_k) \right). \label{eq:61}
\end{equation}
Since $\phi(t)$ is a strictly monotonically increasing function, its inverse function $\phi^{-1}(t)$ exists and  therefore 
\begin{equation}
    \tilde{l}_k=\phi^{-1}(\mu_k), \ \tilde{u}_k=\phi^{-1}(\nu_k). 
\end{equation}
Using the above notation, $\rho$ can be written as:
\begin{align}
    \rho&=\sum_{k=1}^A \frac{\left[ \phi^{'}(\tilde{u}_k)- \phi^{'}(\tilde{l}_k)\right]^2}{\phi(\tilde{u}_k)-\phi(\tilde{l}_k)} \nonumber \\
    &=\sum_{k=1}^A \left[ \frac{\phi^{'}\left(\phi^{-1}(\nu_k)\right)-\phi^{'}\left(\phi^{-1}(\mu_k)\right)}{\nu_k-\mu_k}\right]^2 \cdot  (\nu_k-\mu_k). \label{eq:58}
\end{align}
Next, we define a function $\psi: [0,1] \to \mathbb{R}$ as follows:
\begin{equation}
    \psi(t)=\phi^{'}\left( \phi^{-1}(t)\right),
\end{equation}
and rewrite \eqref{eq:58} as
\begin{equation}
    \rho=\sum_{k=1}^A\left[ \frac{\psi(\nu_k) -\psi(\mu_k)}{\nu_k-\mu_k} \right]^2 \cdot (\nu_k-\mu_k).
\end{equation}
Making use of Lagrange mean value theorem yields:
\begin{align}
    \rho&=\sum_{k=1}^A [\psi^{'}(\zeta_k)]^2 \cdot (\nu_k-\mu_k), \quad \zeta_k \in (\mu_k,\nu_k) \nonumber \\
    & \to \int_{0}^1 [\psi^{'}(t)]^2 \ dt, \quad ({\rm as }\ A \to \infty \ {\rm and}\ \max_{1\leq k \leq A} \lbrace \nu_k-\mu_k\rbrace \rightarrow 0).  \label{eq:61}
\end{align}
To  evaluate the integral above, we  use the following facts:
\begin{equation}
    \left(\phi^{-1}\right)^{'}(t)=\frac{1}{\phi^{'}\left(\phi^{-1}(t)\right)}, \label{eq:phi61}
\end{equation}
and 
\begin{align}
    \psi^{'}(t)&=\phi^{''}\left( \phi^{-1}(t)\right)\cdot (\phi^{-1})'(t) \quad ({\rm the \ chain \ rule}) \nonumber \\
    &=\frac{\phi^{''}\left(\phi^{-1}(t)\right)}{\phi^{'}\left(\phi^{-1}(t)\right)}. \label{eq:62}
\end{align}
Consider the following change of variable:
\begin{equation}
    x=\phi^{-1}(t) \Leftrightarrow t=\phi(x)  \label{eq:xx}
\end{equation}
(hence  $t|_{0}^1 \Rightarrow x|_{-\infty}^{\infty}$ and $dt=\phi^{'}(x)dx$).  Using \eqref{eq:62} we can rewrite  \eqref{eq:61} as:
\begin{equation}
    \rho \rightarrow \int_{-\infty}^{\infty} \left[ \frac{\phi^{''}(x)}{\phi^{'}(x)} \right]^2\phi^{'}(x)dx=\int_{-\infty}^{\infty} \frac{[\phi^{''}(x)]^2}{\phi^{'}(x)}dx. \label{eq:reuse}
\end{equation}
Making use of the result in \eqref{eq:26}, we get  
\begin{equation}
    \rho \to   \frac{1}{\sqrt{2\pi}}\int_{-\infty}^{\infty} x^2 e^{-\frac{x^2}{2}}dx=1,
\end{equation}
and the proof of \eqref{eq:28} is finished. 
\section*{B. \upshape Extensions to general distributions} \label{appendix:A}
The FIM formula in \eqref{eq:crb},
\begin{equation}
       \vec{J}=\sum_{n=1}^N  \left[ \sum_{k=1}^{A} \frac{\left[ \phi^{'}(u_k-s_n)-\phi^{'}(l_k-s_n)\right]^2}{\phi(u_k-s_n)-\phi(l_k-s_n)}  \right]\frac{\partial s_n}{\partial \vec{\theta}} \frac{\partial s_n}{\partial \vec{\theta}^T}, 
\end{equation}
where now $\phi(x)$ denotes a general (differentiable)  cdf, holds for any distribution of noise (indeed, only the specific expression for FIM in \eqref{eq:J} relies on the normal noise assumption; all the other calculations in the proof of \eqref{eq:crb} are valid for a general cdf).
\par The inequality in \eqref{eq:17}, namely
\begin{equation}
   \tilde{\vec{J}} \geq \vec{J} 
\end{equation}
also holds in general as its derivation did not rely on any distributional assumption.
\par The standard CRB (for unquantized data), which generalizes \eqref{eq:J0} to an arbitrary distribution (with zero mean and finite variance), is given by (see, e.g., \cite{stoica2011gaussian}):
\begin{equation}
    \vec{J}_0=\rho_0 \sum_{n=1}^N \frac{\partial s_n}{\partial \vec{\theta}} \frac{\partial s_n}{\partial \vec{\theta}^T}
\end{equation}
where 
\begin{equation}
    \rho_0=\int_{-\infty}^{\infty} \frac{[\phi^{''}(x)]^2}{\phi^{'}(x)}\ dx \label{eq:rho0}
\end{equation}
and where it was implicitly assumed that $\phi^{'}(x)>0$ for any $|x|< \infty$. The result that $\rho \leq \rho_0$, and therefore
\begin{equation}
    \vec{J} \leq \vec{J}_0
\end{equation}
also holds in the present general case, as can be seen directly from \eqref{eq:26}. An interesting fact in this context is that $\rho_0$ in \eqref{eq:rho0} is lower bounded by 1 (see \cite{stoica2011gaussian}), which means that the normal distribution has the smallest matrix $\vec{J}_0$ (in the order of PSD matrices)  in the class of standard FIMs. Note that this is not true for quantized data: for a given set of intervals $\lbrace I_k \rbrace$ the matrix $\vec{J}$ associated with the normal distribution is not necessarily the smallest FIM (indeed we have verified numerically that the FIM for the Laplace distribution, for example, may be smaller for some intervals $\lbrace I_k\rbrace$).
\par Finally we show that the convergence result \eqref{eq:28}, or equivalently \eqref{eq:converg}, continues to hold in the general case. To that end we use the result in  \eqref{eq:reuse}, whose derivation did not rely on the normal distribution assumption,
\begin{equation}
    \rho \rightarrow \int_{-\infty}^{\infty} \frac{[\phi^{''}(x)]^2}{\phi^{'}(x)}dx=\rho_0
\end{equation}
which proves that $\vec{J}$ converges to $\vec{J}_0$.

\bibliographystyle{IEEEtran} 
\normalsize
\bibliography{main}

\begin{thebibliography}{10}
\providecommand{\url}[1]{#1}
\csname url@samestyle\endcsname
\providecommand{\newblock}{\relax}
\providecommand{\bibinfo}[2]{#2}
\providecommand{\BIBentrySTDinterwordspacing}{\spaceskip=0pt\relax}
\providecommand{\BIBentryALTinterwordstretchfactor}{4}
\providecommand{\BIBentryALTinterwordspacing}{\spaceskip=\fontdimen2\font plus
\BIBentryALTinterwordstretchfactor\fontdimen3\font minus
  \fontdimen4\font\relax}
\providecommand{\BIBforeignlanguage}[2]{{%
\expandafter\ifx\csname l@#1\endcsname\relax
\typeout{** WARNING: IEEEtran.bst: No hyphenation pattern has been}%
\typeout{** loaded for the language `#1'. Using the pattern for}%
\typeout{** the default language instead.}%
\else
\language=\csname l@#1\endcsname
\fi
#2}}
\providecommand{\BIBdecl}{\relax}
\BIBdecl

\bibitem{mo2017channel}
J.~Mo, P.~Schniter, and R.~W. Heath, ``Channel estimation in broadband
  millimeter wave {MIMO} systems with few-bit {ADCs},'' \emph{IEEE Trans.
  Signal Process.}, vol.~66, no.~5, pp. 1141--1154, 2017.

\bibitem{chaoyi}
C.-Y. Wu, J.~Li, and T.~F. Wong, ``A {Cram{\' e}r-Rao} bound analysis for
  {mmWave} {PMCW MIMO} radar with quantized observations,'' in \emph{54th
  Asilomar Conf. Signals, Syst. Comput.}, Pacific Grove, USA, Nov. 2020.

\bibitem{sun2020mimo}
S.~Sun, A.~P. Petropulu, and H.~V. Poor, ``{MIMO} radar for advanced
  driver-assistance systems and autonomous driving: {Advantages} and
  challenges,'' \emph{IEEE Signal Process. Mag.}, vol.~37, no.~4, pp. 98--117,
  2020.

\bibitem{mezghani2010multiple}
A.~Mezghani, F.~Antreich, and J.~A. Nossek, ``Multiple parameter estimation
  with quantized channel output,'' in \emph{International ITG Workshop on Smart
  Antennas (WSA)}, Bremen, Germany, Feb 2010.

\bibitem{host2000effects}
A.~Host-Madsen and P.~Handel, ``Effects of sampling and quantization on
  single-tone frequency estimation,'' \emph{IEEE Trans. Signal Process.},
  vol.~48, no.~3, pp. 650--662, 2000.

\bibitem{gianelli2016one}
C.~Gianelli, L.~Xu, J.~Li, and P.~Stoica, ``One-bit compressive sampling with
  time-varying thresholds: {Maximum} likelihood and the {Cram{\'e}r-Rao}
  bound,'' in \emph{Proc. 50th Asilomar Conf. Signals, Syst. Comput.}, Pacific
  Grove, USA, Nov. 2016.

\bibitem{gianelli2019one}
C.~D. Gianelli, ``One-bit compressive sampling with time-varying thresholds:
  The {Cram{\'e}r-Rao} bound, maximum likelihood, and sparse estimation,''
  Ph.D. dissertation, University of Florida, 2019.

\bibitem{kay1993fundamentals}
S.~M. Kay, \emph{Fundamentals of statistical signal processing}.\hskip 1em plus
  0.5em minus 0.4em\relax Prentice Hall PTR, 1993.

\bibitem{stoica1997}
P.~Stoica and R.~L. Moses, \emph{Spectral Analysis of Signals}.\hskip 1em plus
  0.5em minus 0.4em\relax Upper Saddle River, NJ: Prentice-Hall, 2005.

\bibitem{interval_opt}
Y.~Cheng, X.~Shang, and P.~Stoica, ``Interval design for signal parameter
  estimation from quantized data,'' 2021 (to be submitted).

\bibitem{rudin1976principles}
W.~Rudin \emph{et~al.}, \emph{Principles of mathematical analysis}.\hskip 1em
  plus 0.5em minus 0.4em\relax McGraw-Hill New York, 1976, vol.~3.

\bibitem{max1960quantizing}
J.~Max, ``Quantizing for minimum distortion,'' \emph{IRE Trans. Inf. Theory},
  vol.~6, no.~1, pp. 7--12, 1960.

\bibitem{stoica2011gaussian}
P.~Stoica and P.~Babu, ``The {Gaussian} data assumption leads to the largest
  {Cram{\'e}r-Rao} bound [lecture notes],'' \emph{IEEE Signal Process. Mag.},
  vol.~28, no.~3, pp. 132--133, 2011.

\end{thebibliography}

\end{document}